\documentclass[twocolumn, prl, letterpaper]{revtex4}% Physical Review B

\usepackage{graphicx}% Include figure files
\usepackage{dcolumn}% Align table columns on decimal point
\usepackage{bm}% bold math

\begin{document}

\preprint{PREPRINT-DRAFT 0}

\title{Multi-symmetry and Multi-band Superconductivity in Filled-skutterudites\\PrOs$_4$Sb$_{12}$ and PrRu$_4$Sb$_{12}$}
% Force line breaks with \\

\author{R.W. Hill$^1$, Shiyan Li$^2$, M.B.Maple$^3$ and Louis Taillefer$^2$}
\affiliation{$^1$Guelph-Waterloo Physics Institute and Department of Physics and Astronomy, University of
Waterloo, Waterloo, Ontario, Canada\\
$^2$CIAR and D\'{e}partement de physique, Universit\'{e} de
Sherbrooke, Sherbrooke, Qu\'{e}bec, Canada\\
$^3$Department of Physics and Institute for Pure and Applied Physical
Sciences, \\University of California, San Diego, 9500 Gilman Drive, La
Jolla, CA 92093, USA}
%Lines break automatically or can be forced with \\
%\author{Shiyan Li, Louis Taillefer}%
 \email{robhill@Uwaterloo.ca}
%\altaffiliation[Also at]{%
%Authors' institution and/or address\\
%This line break forced with \textbackslash\textbackslash
%}%

%\author{M.B.Maple}
%\affiliation{Department of Physics and Institute for Pure and Applied
%Sciences, \\University of California, San Diego, 9500 Gilman Drive, La
%Jolla, CA 92093, USA}

%\author{Shiyan Li, Louis Taillefer}
%\affiliation{CIAR and D\'{e}partement de physique, Universit\'{e} de
%Sherbrooke, Sherbrooke, Qu\'{e}bec, Canada}
% \homepage{http://www.Second.institution.edu/~Charlie.Author}
%\affiliation{
%Second institution and/or address\\
%This line break forced% with \\
%}%

\date{\today}% It is always \today, today,
             %  but any date may be explicitly specified

\begin{abstract}
Thermal conductivity measurements were performed on
single crystal samples of the superconducting filled skutterudite compounds
PrOs$_4$Sb$_{12}$ and PrRu$_4$Sb$_{12}$ both as a function of temperature and
magnetic field applied perpendicular to the heat current.  In zero
magnetic field, the low temperature electronic thermal conductivity of PrRu$_4$Sb$_{12}$ is vanishingly small, consistent with a fully-gapped Fermi surface.  For PrOs$_4$Sb$_{12}$, however, we find clear evidence for residual electronic conduction
as the temperature tends to zero Kelvin which is consistent with the
presence of nodes in the superconducting energy gap.  The field dependence of
the electronic conductivity for both compounds shows a rapid rise immediately above
H$_{c1}$ and significant structure over the entire vortex state.  In the fully gapped superconductor PrRu$_4$Sb$_{12}$, this is interpreted in terms of multi-band effects.  In PrOs$_4$Sb$_{12}$, we consider the Doppler shift of nodal quasiparticles at low fields and multiband effects at higher fields.
%This is consistent with a semi-classical theory based on a Doppler-shift
%of the quasiparticle spectrum through coupling to the superfluid flow
%around magnetic vortices.
\end{abstract}

%\pacs{Valid PACS appear here}% PACS, the Physics and Astronomy
                             % Classification Scheme.
%\keywords{Suggested keywords}%Use showkeys class option if keyword
                              %display desired
\maketitle

%\section{Introduction}
Fundamental research into superconductivity is extremely important because of the potentially revolutionary technology were this phenomena available for widespread commercial use.  With improvements in materials science,
new materials that are superconducting continue to be discovered and
experimental studies on them
shed light on where our knowledge about superconductivity can be enhanced.  The filled
skutterudite family of materials is one such example, exhibiting many
features that suggest in some variants its superconductivity may be conventional,
whilst in others unconventional.  A comparison between the superconducting properties of
each type of material is certainly instructive.

Some of the more unusual properties have been seen in the heavy-fermion superconductor PrOs$_4$Sb$_{12}$.
For example, in angle-resolved magneto-thermal
conductivity experiments \cite{Izawa-PRL},  a change in the symmetry of the small anisotropy in the
conductivity is
interpreted as evidence for a multiphase superconducting phase-diagram.  In this case,
switching between a phase that has two nodes in
the superconducting
order parameter to one which has four nodes, as the magnetic field is increased.  More recently, the thermal
conductivity has been measured to even lower temperatures \cite{Seyfarth1, Seyfarth2}.
Surprisingly, the measurements show an absence of any residual
electronic conduction in zero field, which is inconsistent with the
earlier measurements in either superconducting phase.  Moreover, the
large magnetic field dependence is interpreted as resulting from
multi-band effects through analogy with well-known multiband
superconductors such as MgB$_2$ \cite{Sologubenko}.

On the other hand, the isostructural but non-heavy fermion compound, PrRu$_4$Sb$_{12}$, appears to be a conventional s-wave superconductor with exponential temperature dependencies observed in specific heat \cite{Takeda, Fredericks} and superfluid density \cite{Chia}.

In this Letter, we present low temperature thermal conductivity
measurements that provide evidence for a fully-gapped order parameter in PrRu$_4$Sb$_{12}$ and a nodal superconducting order parameter in PrOs$_4$Sb$_{12}$.  We also demonstrate that the field dependence of the fully-gapped superconductor is consistent with multiband superconductivity. In the nodal superconductor, the field dependence immediately above $H_{c1}$ is consistent with being due
to the Doppler shift of nodal quasiparticles, with additional multiband effects occuring at fields above $H_{c2}/2$.  Finally, we explore the possibility that multiband superconductivity is a generic feature of filled-skutterudite superconductors.

%\section{Experimental Information}
The single crystal samples were oriented using Lau\'{e} x-ray
backscattering.  In the case of PrOs$_4$Sb$_{12}$, the sample
was then polished to a cuboid of dimension (600 x 100 x 100) $\mu$m.  For PrRu$_4$Sb$_{12}$ the sample used was as-grown with dimensions (2500 x 188 x 380) $\mu$m.
In both cases, four silver wires were then attached using high purity indium solder.

The thermal conductivity $\kappa$ was measured using a single heater-two
thermometer method.  The heat current was supplied along the
$a$-axis
direction and the magnetic field applied perpendicular to this.
The
measurements were made in a dilution refrigerator by varying the
temperature from 0.04~K to $> 0.7$~K at fixed magnetic field.  The
samples
were field-cooled by cycling to $T>2$~K before changing the field.
The
error in the absolute value of the conductivity is estimated to be
approximately $10\%$.  The relative error between temperature sweeps at
different fields is of order 1\%.

%\section{Results and Discussion}

%Outline:
%(i) Zero Field: sizeable linear term in POS, compare with d-wave estimates. vanishingly small value in PRS (again compare with estimates: vF - larger, v2 - smaller, therefore k0/T even bigger).  Consistent with other measurements that POS is nodal, PRS is fully gapped.  Since see nodal effect in one but not in other, rules out contact effect.
%(ii) Field Dependence: Nodal has Volovik and Zeeman shift, while fully gapped only has Zeeman shift.  POS is quantitatively consistent with Volovik.  PRS must be multiband scenario.  Likelihood is both are multiband, one is nodal and the other fully gapped.

\begin{figure}
\resizebox{\columnwidth}{!}{
\includegraphics{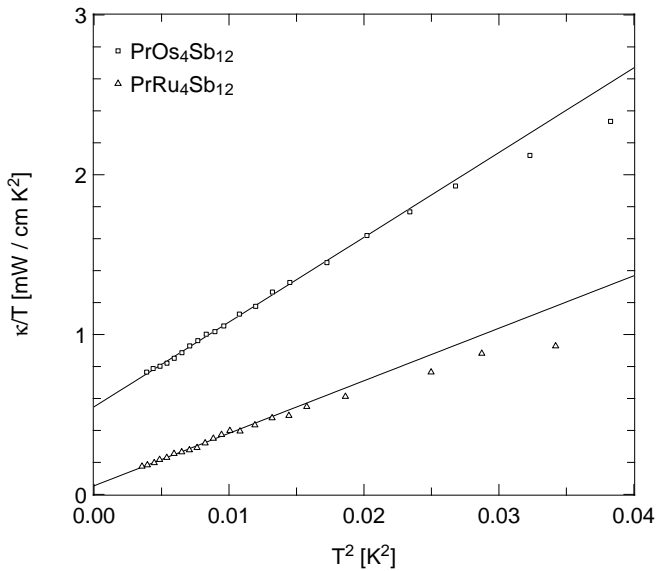}}
\caption{\label{fig:rlt} Thermal conductivity divided by
temperature $T$ versus $T^2$ in zero field for PrOs$_4$Sb$_{12}$ and PrRu$_4$Sb$_{12}$.  The lines are linear fits to the low temperature data ($T < 0.15$ K) which are then extrapolated to the $T = 0$ axis.}
\end{figure}

\noindent {\it Zero Magnetic Field}:  In Fig.~\ref{fig:rlt}, the temperature dependence of the zero-field thermal conductivity is plotted for both materials.  Since the measured quantity is the total thermal conductivity, it is necessary to separate the contributions from electrons and phonons.  Using the well-established model based on kinetic theory (see, for example, \cite{Taillefer}), in the low temperature limit, the electronic contribution is linear in temperature, while phonons are cubic in temperature.  Fitting the measured conductivity ($\kappa$) to the form:
\begin{equation}
\frac{\kappa}{T} = \frac{\kappa_{0}}{T} + \beta T^2
\label{eq:rho}
\end{equation}
and extrapolating the fit to zero temperature we obtain a value for the residual electronic conductivity divided by $T$, $\kappa_0 / T$.  The coefficient $\beta$ in the above expression represents the phonon contribution in this simple analysis.  Fitting the data below $T=0.15$, the values obtained for each material are $0.46 \pm 0.07$ mW/K$^2$cm and $0.058 \pm 0.007$ mW/K$^2$cm for PrOs$_4$Sb$_{12}$ and PrRu$_4$Sb$_{12}$ respectively.

%Fig 1 - residual conductivity in POS and PRS

In PrRu$_4$Sb$_{12}$, the value of $\kappa_0 / T$ is an order of magnitude smaller than that measured in PrOs$_4$Sb$_{12}$. In instances where the extrapolated value is very small, the simplicity of the model used to separate electrons and phonons is exposed.  Improved fitting accuracy has been demonstrated using an empirically determined variable power law for the phonon contribution \cite{Sutherland}.  In such cases the extrapolated value for the residual linear term is always found to be reduced and in this case we find $\kappa_0/T = 8 \pm 17 \mu$W/K$^2$cm and a phonon exponent of T$^{1.6}$. Within error bars such a value can be considered as zero indicating the absence of any residual electronic conductivity and therefore consistent with a fully-gapped superconducting state.

In contrast, the observation of a significant finite value for the zero-temperature extrapolation of the linear electronic
conductivity in PrOs$_4$Sb$_{12}$ is incontrovertible evidence for nodes on the Fermi surface of the superconducting
order parameter.

As a consistency check, a calculation of the magnitude of the residual linear term assuming a simple $d$-wave symmetry of the superconducting gap \cite{Graf} and weak coupling can be made for both systems.  In PrOs$_4$Sb$_{12}$, we estimate $\kappa_{0}/T = 0.3$ mW/K$^2$cm.
However, we note that such a 4-fold symmetry would only be consistent with the high-field superconducting phase observed
by angle-resolved magneto-thermal conductivity measurements \cite{Izawa-PRL}.  For PrRu$_4$Sb$_{12}$, this simple calculation would give an even larger value because of the smaller gap magnitude (assuming a similar Fermi velocity), which is consistent with very small extrapolated value being interpreted as evidence for a fully gapped Fermi surface.

In spite of the more than order of magnitude difference in the electronic conductivity, the apparent phonon contribution in each material is very similar, as might be expected given the identical structure of these materials.

The observation of a linear term in PrOs$_4$Sb$_{12}$ in this study directly contradicts the observation of other low temperature thermal conductivity measurements where the data extrapolates to zero with a rapid temperature dependence that exceeds $T^3$ \cite{Seyfarth1,Seyfarth2}.  Rapid temperature dependencies such as these are known to occur when electron-phonon decoupling hampers the measurement of the intrinsic electronic conductivity \cite{Smith} and is thought to occur in other superconducting systems \cite{Hill-Nature, Nakamae}.  In CeCoIn$_5$ in particular, early measurements indicated a large $T^{3.4}$ temperature dependence and zero electronic residual conductivity \cite{Movshovich}.  More recent measurements show a large residual electronic conductivity and a lower $T^2$ temperature dependence \cite{Tanatar}.  In the study reported here, the use of an identical technique to measure each material with identical contacts and yet dramatically different but nevertheless consistent results is compelling.

\begin{figure}
\centering
\resizebox{\columnwidth}{!}{
\includegraphics{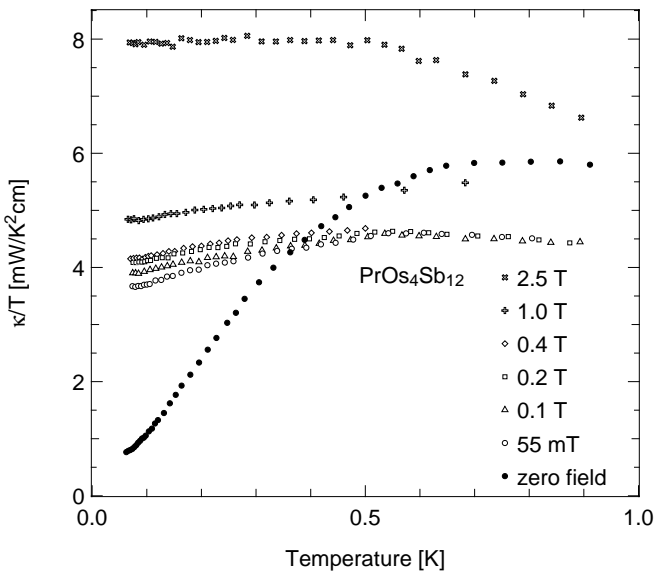}}% Here is how to import EPS art
\caption{
\label{fig:POS_B} Thermal
conductivity divided by
temperature $T$ versus $T$ for
PrOs$_4$Sb$_{12}$ at different magnetic fields.  The heat current is
applied along the $a$-axis of the single
crystal sample. The magnetic field is applied perpendicular to this direction.}
\end{figure}

\begin{figure}
\centering
\resizebox{\columnwidth}{!}{
\includegraphics{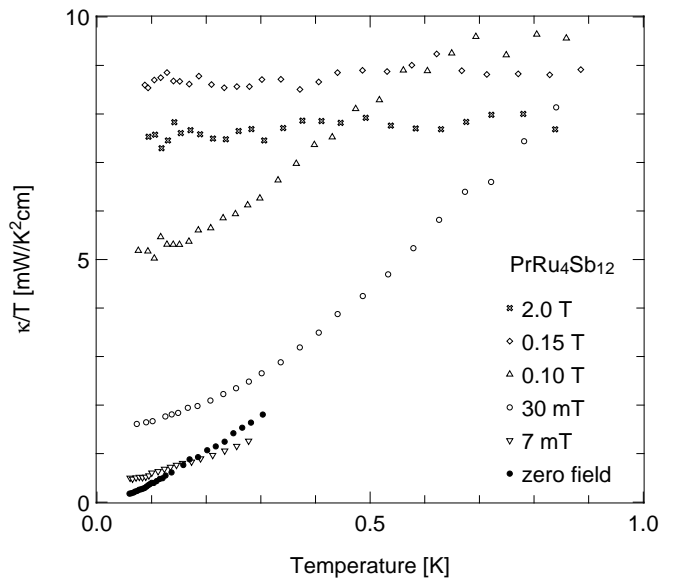}}% Here is how to import EPS art
\caption{ \label{fig:PRS_B} Thermal
conductivity divided by
temperature $T$ versus $T$ for
PrRu$_4$Sb$_{12}$ at different magnetic fields.  The heat current is
applied along the $a$-axis of the single
crystal sample. The magnetic field is applied perpendicular to this direction.}
\end{figure}

\noindent {\it Magnetic Field Dependence}:  The temperature dependence of the thermal conductivity for different magnetic fields applied perpendicular to the direction of the heat current is shown in Fig.~\ref{fig:POS_B} for PrOs$_4$Sb$_{12}$ and Fig.~\ref{fig:PRS_B} for PrRu$_4$Sb$_{12}$.

%Fig 2 - field dependence for POS

The data for both materials show features that are qualitatively similar, although there are some distinct differences.  In zero field, the data have the steepest temperature dependence, which in PrOs$_4$Sb$_{12}$ rises to a plateau above $T$ = 0.6 K.  Measurements at higher temperature show this to be a peak \cite{Rahimi}.  In both materials, as soon as a magnetic field is applied the $T$-dependence is suppressed by a dramatic increase in $\kappa/T$ at the lowest temperatures.  For subsequent higher fields, the curves at low temperatures then remain approximately parallel as the field increases by orders of magnitude and the system enters its field-induced normal state.  We note in passing that the data for PrOs$_4$Sb$_{12}$ at low fields show a striking similarity to the behaviour observed in ultra-pure YBa$_2$Cu$_3$O$_7$ \cite{Hill-PRL1}.

\begin{figure}
\centering
\resizebox{\columnwidth}{!}{
\includegraphics{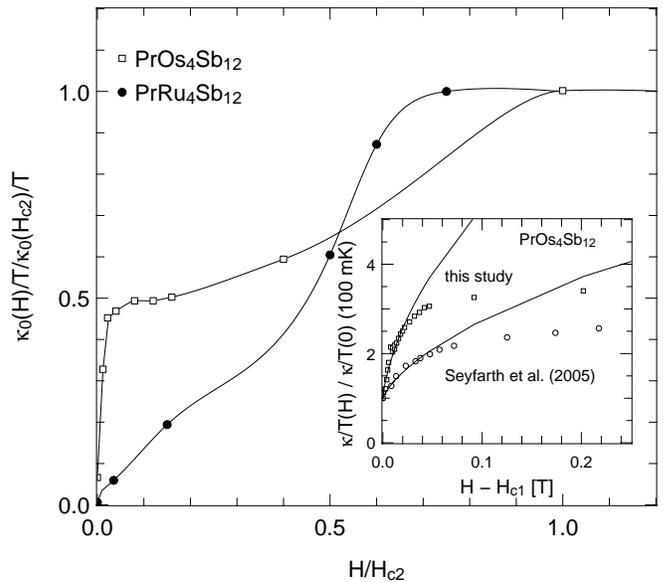}}% Here is how to import EPS art
\caption{\label{fig:k_B} Magnetic field dependence of the extrapolated zero temperature thermal conductivity of PrOs$_4$Sb$_{12}$ and PrRu$_4$Sb$_{12}$.  The lines are guides to the eye in each case.  The inset shows the field dependence of the conductivity normalised to the zero-field valued for PrOs$_4$Sb$_{12}$ both from this study and from earlier work\cite{Seyfarth1}.  The lines result from fitting to a semi-classical theory \cite{Kubert} based on a Doppler shift of the nodal quasiparticle spectrum.}
\end{figure}

In order to explore the magnetic field dependence more clearly, the value of the temperature dependence extrapolated to $T=0$ K at each field is extracted, normalised to the normal state value, and then plotted in Fig.~\ref{fig:k_B} as a function of magnetic field normalised to the upper critical field ($H_{c2}$).  Values used for the upper critical field are $H_{c2}$ = 2.0 T and 0.2 T for PrOs$_4$Sb$_{12}$ and PrRu$_4$Sb$_{12}$, respectively.  In the simplest fully gapped superconductor, the field dependence of the residual electronic conduction rises exponentially with magnetic field on a scale set by $H_{c2}$, as has been observed experimentally in V$_3$Si \cite{Boaknin}. In contrast,  Fig.~\ref{fig:k_B} shows the field dependence for both materials rises rapidly at low fields.  On this reduced magnetic field scale, the increase is considerably more rapid in PrOs$_4$Sb$_{12}$ than in PrRu$_4$Sb$_{12}$.  Futhermore, over the entire field range both materials show appreciable structure.   Such behaviour is reminiscent of that seen in multiband superconductors such as MgB$_2$ \cite{Sologubenko} and NbSe$_2$ \cite{Boaknin} and has been suggested already for PrOs$_4$Sb$_{12}$ \cite{Seyfarth1},.  Since the zero field behaviour has already established the fully gapped nature of the superconducting order parameter in PrRu$_4$Sb$_{12}$, this is the most likely explanation in this material.

%Fig 3 - field dependence for PRS

In PrOs$_4$Sb$_{12}$, the picture is less clear because of the presence of nodal quasiparticles.  The principal effect of magnetic field on superconductors with a nodal order parameter is the Doppler shift of the nodal quasiparticle energy spectrum through coupling to the superfluid flow around magnetic vortices.  This leads to an increase in the density of states and associated properties.  The original semi-classical treatment by Volovik \cite{Volovik} has been extended to explain the effect on thermal transport by Hirschfeld and Kubert \cite{Kubert} with the increase dependent on the normal state scattering rate and given by
\begin{equation}
\frac{\kappa(0,H)}{T} = \frac{\kappa_0}{T}\left[\frac{\rho^2}{\rho\sqrt{1+\rho^2} - \sinh^{-1}\rho}\right]
\label{eq:rho}
\end{equation}
%{\it equation for rho}\
The parameter $\rho = \sqrt{3\Phi_0}\gamma/a\hbar v_F\sqrt{H}$ is related to the impurity bandwidth ($\gamma$) and the normal state scattering rate ($\Gamma$), $\gamma \sim \Gamma^2$.  In the inset to Fig~\ref{fig:k_B}, we plot the normalised increase in thermal conductivity at $T=100$ mK with magnetic field for both the data presented here and for the earlier measurements \cite{Seyfarth1}. In each case, the initial increase is fitted to extract values for $\rho$. The values obtained are $\rho=90$ for this study and $\rho=183$ for the data from \cite{Seyfarth1}.  We note that there is some variability in exactly what range of data to fit since at higher fields there is a crossover to a field independent region.  We have chosen to fit the data up to 80\% of the plateau value conductivity.  Adjusting the range in each case causes a shift in the values of $\rho$ by approximately 10\%.  Nonetheless, the important point is that the values for $\rho$ in each case differ by a factor of two, which implies that the normal state scattering rates and, hence, normal state conductivity should vary by a factor of four.  Comparing our normal state conductivity value, $\kappa_0(H = 2 \text{T})/T = 8$ mW/K$^2$cm, with that from Ref\cite{Seyfarth1}, $\kappa_0(H = 2 \text{T})/T = 2$ mW/K$^2$cm, we see that quantitatively this is exactly the case.  The rapid increase with magnetic field therefore scales with the normal state scattering rate in a manner that is entirely predicted by a Doppler shift of the quasiparticle energy spectrum.  Unfortunately, and as has been established already from similar measurements in cuprates \cite{Hill-PRL1}, the absolute value of the normal state scattering rate obtained from these fits is not in agreement with that from the normal state transport properties.  This most likely reflects the limitations of the current theory.

At higher fields in PrOs$_4$Sb$_{12}$, the conductivity exhibits a plateau and then increases in a super-linear fashion up to the normal state value at $H_{c2}$.  This aspect is again reminiscent of the behaviour seen in multiband superconductors.  Consequently, we postulate that this high-field dependence results from quasiparticles associated with a second, larger superconducting gap covering a separate Fermi surface sheet.  The resulting picture for PrOs$_4$Sb$_{12}$ is that it has one band with a nodal order parameter and another distinct band that is fully gapped.  This scenario is appealing for a number of reasons.  First, if we postulate that the gap maximum on the nodal band is somewhat smaller than the gap maximum on the fully gapped band, then our estimate for the value of the residual electronic conductivity would be revised upwards and closer to the value we measure.  Second, the rapid rise of the conductivity at low fields would now be considered on a scale set by a presumably lower critical field for the band with a nodal gap.  This may provide a more reasonable absolute value for the normal state scattering rate. Third, it may provide a consistent picture of the angle resolved magneto-thermal conductivity measurements \cite{Izawa-PRL}. If we postulate a degree of anisotropy to the fully gapped system with a minimum orthogonal to the nodes of the other gap, then high field or high temperature measurements would pick up a four-fold symmetry due to the nodes on the gap of one band and the anisotropy on the fully-gapped other band. At low fields and temperatures, only the anisotropy due to the nodal sheet would survive with quasiparticle excitations on the (anisotropic) fully-gapped band exponentially suppressed.  Such a system would thus have a superconducting state that is not only multi-band, but also multi-symmetric.  The reason behind why one order parameter should have nodes and the other fully-gapped may be related to the heavy-fermion character of particular sheets of the Fermi surface.

%\section{Conclusion and Summary}

In conclusion, we have measured the low temperature thermal conductivity of the filled skutterudites PrRu$_4$Sb$_{12}$ and PrOs$_4$Sb$_{12}$ as a function of transverse magnetic field. Measurements in zero field indicate that in PrRu$_4$Sb$_{12}$, there is no evidence of any residual conduction, consistent with a fully gapped superconducting state.  Conversely, the extrapolated data in PrOs$_4$Sb$_{12}$ shows a distinct linear contribution consistent with low energy electronic excitation and therefore with a superconducting gap with nodes on the Fermi surface. A rapid increase of the conductivity with magnetic field is observed in both materials.  In PrRu$_4$Sb$_{12}$, the increase in conductivity is less rapid but still above what one would expect for a simple isotropic gapped superconductor.  This increase is attributed to multiband superconductivity.  In PrOs$_4$Sb$_{12}$, the increase is consistent with nodal quasiparticle excitations coupling to the superfluid flow around magnetic vortices.  Moreover the magnitude of this increase is found to scale appropriately with the normal state scattering rate when compared to existing data for measurements on samples with shorter quasiparticle lifetimes.  At higher fields, the superlinear increase in conductivity is attributed to quasiparticle excitations associated with a large fully-gapped band on a separate sheet of the Fermi-surface.  If such a scenario is correct, it is possible that multiband superconductivity may be a generic feature of superconducting filled skutterudite systems.  As this study demonstrates, this appears to be the case for both PrRu$_4$Sb$_{12}$ and PrOs$_4$Sb$_{12}$.

\begin{acknowledgments}
This work was funded by NSERC of Canada.  SL and LT are supported by the Canadian Institute For Advanced Research.
Research at UCSD was supported by the US Department of Energy under Grant No. DE FG02-04ER46105.
%We wish to acknowledge the support of the author community in using
%REV\TeX{}, offering suggestions and encouragement, testing new
%versions,
%\dots.
\end{acknowledgments}

%\bibliography{Skutterudite_preprint1}

%\appendix

%\section{Appendixes}

\end{document}